\documentclass[a4paper, aps, prl, twocolumn, longbibliography]{revtex4-1}
\usepackage{amsmath, amssymb}
\usepackage{graphicx}
\usepackage{comment}
\usepackage{hyperref}
\usepackage{soul,color}
\usepackage{braket}
\usepackage{cleveref}

\graphicspath{{./graphs/===}}

\let\Re\relax \DeclareMathOperator\Re{\mathrm{Re}}
\DeclareMathOperator\tr{\mathrm{tr}}%

\newcommand{\average}[1]{\left\langle{#1}\right\rangle}

\begin{document}

\title{Environment-Induced Rabi Oscillations in the Optomechanical Boson-Boson Model}

\author{Y. Minoguchi}
\affiliation{Vienna Center for Quantum Science and Technology, Atominstitut, TU Wien, 1040 Vienna, Austria}
\author{P. Kirton}
\affiliation{Vienna Center for Quantum Science and Technology, Atominstitut, TU Wien, 1040 Vienna, Austria}
\author{P. Rabl}
\affiliation{Vienna Center for Quantum Science and Technology, Atominstitut, TU Wien, 1040 Vienna, Austria}

\date{\today}

\begin{abstract}
We analyze the strong-coupling dynamics of a driven harmonic oscillator whose energy is modulated by a continuum  of other bosonic modes. 
This type of system-bath interaction appears, for example, in optomechanical or equivalent circuit QED setups, where the frequency of a confined photonic mode depends linearly on  a fluctuating boundary.
Compared to the canonical spin-boson model, where coupling to bath modes only leads to decoherence, the role of the environment in such systems is more complex, since it also provides the only source of nonlinearity.
We show that even for an unstructured bath, these environment-induced nonlinearities can dominate over decoherence processes resulting in Rabi oscillations and the formation of highly non-classical states. 
These findings provide important insights into the non-Markovian dynamics of higher-dimensional  open quantum systems and for realizing few-photon optical nonlinearities through strong interactions with a bath.
\end{abstract}

\maketitle

Quantum systems are never fully isolated from their surroundings. Thus, to accurately model quantum phenomena and their applications it is key to precisely understand the influence of the environment on the system dynamics. In many situations the coupling to the environment is sufficiently weak that it can be described, for example, in terms of a master equation for the reduced system density operator~\cite{Breuer_book}.
For stronger coupling or non-Markovian baths, the dynamics becomes considerably more involved and general methods to treat such scenarios are no longer available~\cite{Breuer_book,weiss_book,Vega2017}.
In this context, the spin-boson model~\cite{Leggett1987}, which describes a single two-level system coupled to a continuum of bosonic modes, has emerged as a prototype for studying quantum dissipation effects. However, while details  depend non-trivially on the coupling strength and spectrum of the bath~\cite{Leggett1987,LeHur10,LEHUR2018451}, the role of the environment in this setting is rather restricted. 
Starting from Rabi oscillations of the isolated spin, increasing the system-environment coupling simply introduces stronger and stronger damping with, eventually, a transition into a classical regime with overdamped~\cite{Toulouse69,Guinea85,Leggett1987} or localized dynamics~\cite{Bray82,Chakravarty82}.

A very different type of system-bath interaction has recently attracted a lot of attention in the field of optomechanics~\cite{Aspelmeyer2014}, where the frequency of an optical resonator mode is modulated by a vibrating end-mirror or other fluctuating boundary. 
In this case, the system of interest, the optical field, is a harmonic oscillator, which by itself does not exhibit any genuine quantum mechanical features. This leads to an interesting situation; coupling to the continuum of bath modes induces decoherence as well as effective nonlinearities necessary to observe non-classical effects. In this Letter we study the dynamics of this \textit{boson-boson model} under weak and strong driving conditions. Specifically, we examine the effects of photon blockade and environment-induced Rabi oscillations, both of which signify the transition from linear classical to nonlinear quantum behavior.
We find that strong nonlinearities can even be induced through coupling to unstructured environments, where simple arguments derived for individual mechanical modes~\cite{Mancini1997, Bose1997, Rabl2011, Nunnenkamp11} are not applicable. We discuss the dependence of these effects on the spectral properties of the bath, i.e., its degree of non-Markovianity, and show that, surprisingly, a significant level of non-classicality can even be observed in the strongly coupled, overdamped regime.

\begin{figure}
    \includegraphics[width=\columnwidth]{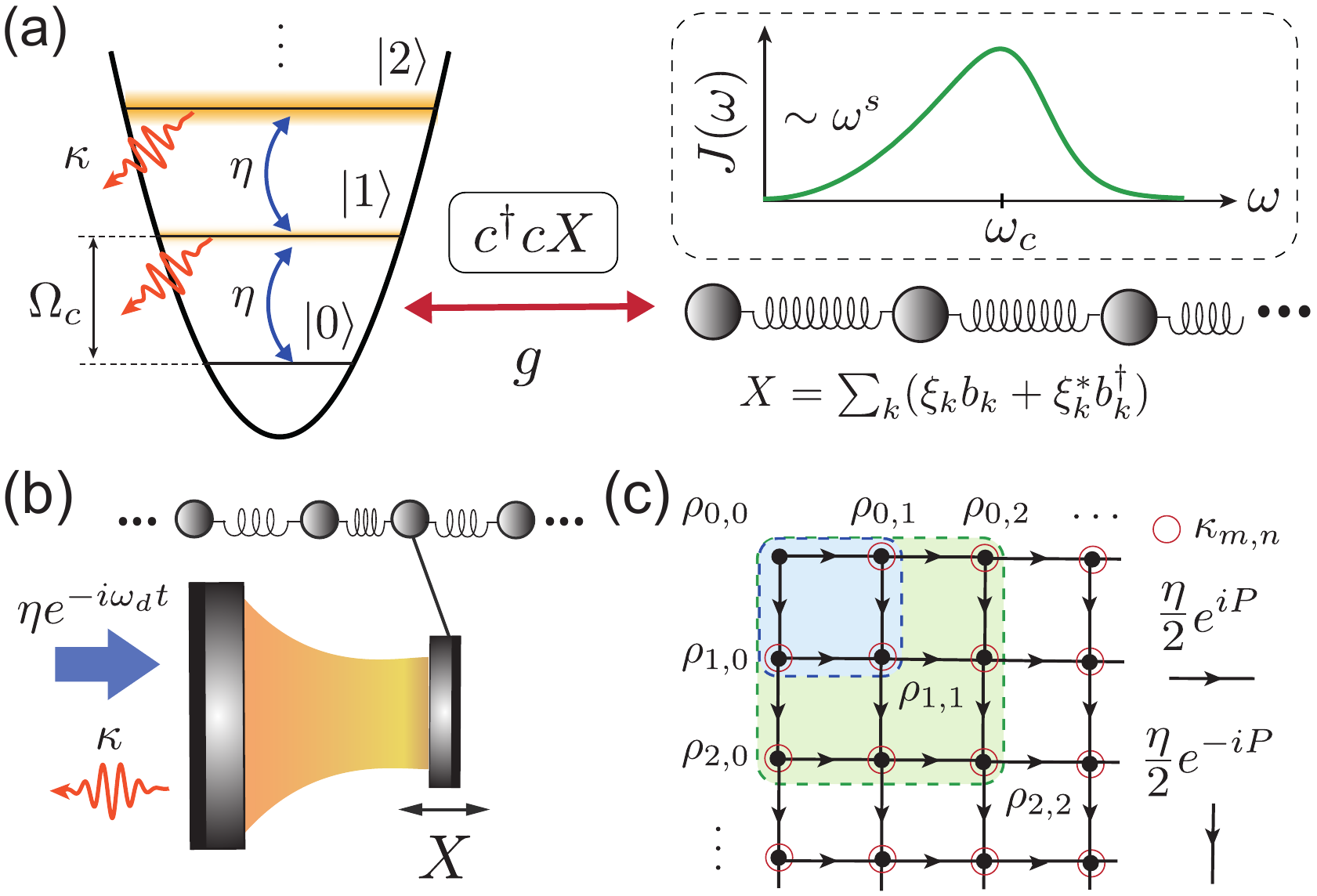}
    \caption{{The boson-boson model}.
    (a) The frequency of a driven oscillator is modulated by the collective coordinate $X$ of a bosonic environment. The bath is characterized by a spectral density $J(\omega)$ with low-frequency exponent $s$ and cutoff $\omega_c$. 
    (b) Optomechanical implementation of the model. 
    The frequency of an optical resonator depends on a moving mechanical boundary, which itself is part of a large phononic reservoir.
    (c) Illustration of the coupled set of equations~\eqref{eq:rhonm} for the reduced density matrices $\rho_{m,n}$, which are used to derive the spectrum $S_{\alpha,\kappa}^{(s)}$ (one-photon sector) and the correlation function $g^{(2)}(0)$ (two-photon sector) in the weak driving limit.}
    \label{fig:Fig1}
\end{figure}

\emph{Model.}---We consider the generic setting depicted  in Fig.~\ref{fig:Fig1}(a), where the energy of a high-frequency cavity mode, with bare frequency $\Omega_c$ and annihilation operator $c$, depends linearly on the collective coordinate $X$ of a low-frequency bosonic bath. 
The cavity is driven by an external field. In the frame rotating with the driving frequency $\omega_d$  the Hamiltonian reads ($\hbar=1$)
\begin{equation}\label{eqn:BBmodel}
H= -\Delta_c c^{\dagger}c +\frac{\eta}{2}(c+c^{\dagger}) + g c^{\dagger}c X + H_{B}.
\end{equation}
Here, $\Delta_c = \omega_d-\Omega_c$ and $\eta$ denote the detuning and strength of the drive and  $H_B=\sum_{k}\omega_k b_k^{\dagger}b_k$ is the bath Hamiltonian with mode annihilation operators $b_k$ and frequencies $\omega_k$. 
For a single bath mode,  $X=(b+b^\dag)$, Eq.~\eqref{eqn:BBmodel} reduces to the standard optomechanical Hamiltonian with coupling strength $g$~\cite{Aspelmeyer2014}.
It describes, for example, a Fabry-Perot cavity with a moving end-mirror as illustrated in Fig.~\ref{fig:Fig1}(b). Similar interactions are found in systems of trapped atoms~\cite{Gupta07,Neumeier18,Shamoon18} or dispersively coupled  $LC$ resonators~\cite{Johansson14, Heikkila2014, Eichler2018} where also much stronger couplings and more complex mode structures can occur. In such general cases $X= \sum_k (\xi_k b_k+ \xi_k^*b_k^\dag)$ and the coupling to the bath is characterized by the spectral density $J(\omega) =g^2  \sum_k\vert \xi_k\vert^2 \delta(\omega-\omega_k)$. 
For the current analysis we follow the usual convention \cite{weiss_book} and take $J(\omega)$ to have the generic form
\begin{equation}
J(\omega)= 2\alpha \omega^s\omega_c^{1-s} e^{-\omega/\omega_c},
\end{equation}
where $\alpha$ is the dimensionless system-bath coupling strength and $\omega_c$ is the cutoff frequency. 
The exponent $s$ determines the low-energy behavior. 
Below we will consider Ohmic, $s=1$, and super-Ohmic, $s=2$, cases. 

To describe realistic conditions we also include the bare decay of the cavity mode with rate $\kappa$. 
In the regime of interest,  $\Omega_c\gg\kappa, \omega_c, k_BT/\hbar$, where $T$ is the temperature, these losses can be modeled by a weak linear  interaction with a Markovian, zero-temperature reservoir, which can be straightforwardly eliminated~\cite{ZollerGardiner04}. 
The resulting dynamics of the full density operator $\rho$ is then given by the master equation  (see~\footnote{See Supplemental Material, which cites Refs.~\cite{makri_makarov_1995_i, makri_makarov_1995_ii, Vagov2011, Strathearn2017}, for discussion of the derivation of the master equation, the formulation of the perturbation theory and details of the numerical methods used.} for more details)
\begin{equation}\label{eq:ME}
\dot\rho = -i[H, \rho] + \frac{\kappa}{2}(2c\rho c^\dag -c^\dag c \rho - \rho c^\dag c).
\end{equation}
 The system is initially in the state $\rho(0) = |0\rangle\langle 0| \otimes \rho_{\mathrm{th}}$, where $\rho_{\mathrm{th}}$ is the thermal equilibrium state of the bath. 
In the absence of the non-Markovian bath, when the driving field is switched on, the cavity  simply evolves into a coherent state 
with a steady-state occupation number $\lim_{t\rightarrow \infty}\average{c^\dag c}_t=\eta^2/ (4\Delta_c^2+\kappa^2)$. 
In the following we are interested in the deviation from this classical behavior when the coupling to the environment is increased.

\emph{Weak Driving Limit.}---We first consider the limit of weak driving, $\eta\rightarrow 0$, where only the lowest photon number states $\ket{n}$ are important and a systematic perturbative expansion in terms of the small parameter $\eta/\kappa$ can be performed~\cite{Rabl2011,Note1}. 
We start with a unitary transformation to the polaron frame, $H_P= \text{e}^{-ic^{\dagger}c P}H\text{e}^{ic^{\dagger}c P}$, with $P = ig\sum_k (\xi^*_kb^\dag_k-\xi_kb_k)/\omega_k$. We obtain 
\begin{equation}
    H_P = -\Delta c^{\dagger}c -\Delta_p c^{\dagger}c^{\dagger}cc +\frac{\eta}{2}(\text{e}^{iP}c+c^{\dagger}\text{e}^{-iP})+H_B,
\end{equation}
where $\Delta=\Delta_c+\Delta_p$ is the renormalized cavity frequency. Here $\Delta_p =\int_0^{\infty}\mathrm{d}\omega J(\omega)/\omega = 2\alpha \omega_c \Gamma(s)$ is the polaron shift, with $\Gamma(x)$ the Gamma function. 

For an undriven cavity ($\eta=0$) the Hamiltonian $H_P$ is diagonal with eigenstates which are cavity photons dressed by displaced bath states, $\ket{\Psi_{n_c,\{ n_k\}}}=\ket{n_c} e^{-in_cP}\ket{ \{ n_k\}}$, and energies $E_{n_c,\{n_k\}}=(\Omega_c-\Delta_p) n_c-\Delta_p n_c(n_c-1)+\sum_k \omega_k n_k$ in the lab frame.
Apart from an overall shift of the cavity frequency, there is also a reservoir-mediated attractive interaction, $\Delta_p n_c(n_c-1)$, between dressed photon states, due to the dependence of the bath displacement on the photon number.
However, the presence of a continuum of bath modes  is inevitably tied to decoherence and it is \textit{a priori} unclear whether this apparent nonlinearity can  be harnessed to produce non-classical states of the bare cavity.  

To evaluate the effect of a small, but finite, driving strength we use the master equation in the polaron frame, Eq.~\eqref{eq:ME}, to derive a hierarchy of coupled equations for the reduced density matrices $\rho_{m,n} = \bra{m} \rho \ket{n}$, 
\begin{multline}\label{eq:rhonm}
\dot{\rho}_{m,n}  = \left(i\Delta_{mn}-\frac{\kappa_{mn}}{2}\right)\rho_{m,n} + i \frac{\eta}{2}\sqrt{n} \rho_{m,n-1}e^{iP(t)} \\
-i\frac{\eta}{2}\sqrt{m}e^{-iP(t)}\rho_{m-1,n} + O(n_0^{m+n+1/2} ),
\end{multline}
where $\kappa_{mn}=\kappa (m+n)$ and $\Delta_{mn} = \Delta_c(m-n) + \Delta_p(m^2-n^2)$. 
As depicted in Fig.~\ref{fig:Fig1}(c), initializing in the state $\rho(0)$, we can iteratively solve these equations.
We may then calculate the relevant expectation values $\average{(c^\dagger)^n c^n}_t = n!\tr_B\{\rho_{n,n}(t)\}+O(n_0^{n+1/2})$, to lowest order in $n_0$.

We first use this approach to discuss the normalized excitation spectrum $S_{\alpha, \kappa}^{(s)}(\Delta_c) = \lim_{t\rightarrow \infty}\average{c^{\dagger}c}_{t}/n_0$, which is given by~
\cite{Mahan65,Rabl2011,Nunnenkamp11,Note1}
\begin{equation} \label{eq:SFormula}
S^{(s)}_{\alpha,\kappa}(\Delta_c) = \frac{\kappa}{2}\Re\int_0^{\infty}\mathrm{d}\tau \, \text{e}^{(i(\Delta_c+\Delta_p)+\kappa/2)\tau} e^{-F_2(\tau)}.
\end{equation}
Here, $e^{-F_2(\tau)} = \average{e^{iP(\tau)}\text{e}^{-iP(0)}}_{\text{th}}$ is the displacement correlation function of the environment, with $F_2(t)  = \int_0^{\infty}\mathrm{d}\omega\,J(\omega) \left(\coth\left( \beta\omega/2\right) (1-\cos(\omega t)) + i\,\sin(\omega t)\right)/\omega^2$ and $\beta=1/k_B T$.
For simplicity we focus on the limit $k_BT/\hbar\omega_c \ll 1$, where in general $F_2(t)=2\alpha \Gamma(s-1)[1-(1+i\omega_c t)^{1-s}]$ and for $s=1$ this reduces to $F_2(t)=2\alpha \ln(1+i\omega_c t)$.
A detailed analysis of finite temperature effects  will be given elsewhere~\cite{unpub}.

\begin{figure}
    \includegraphics[width=\columnwidth]{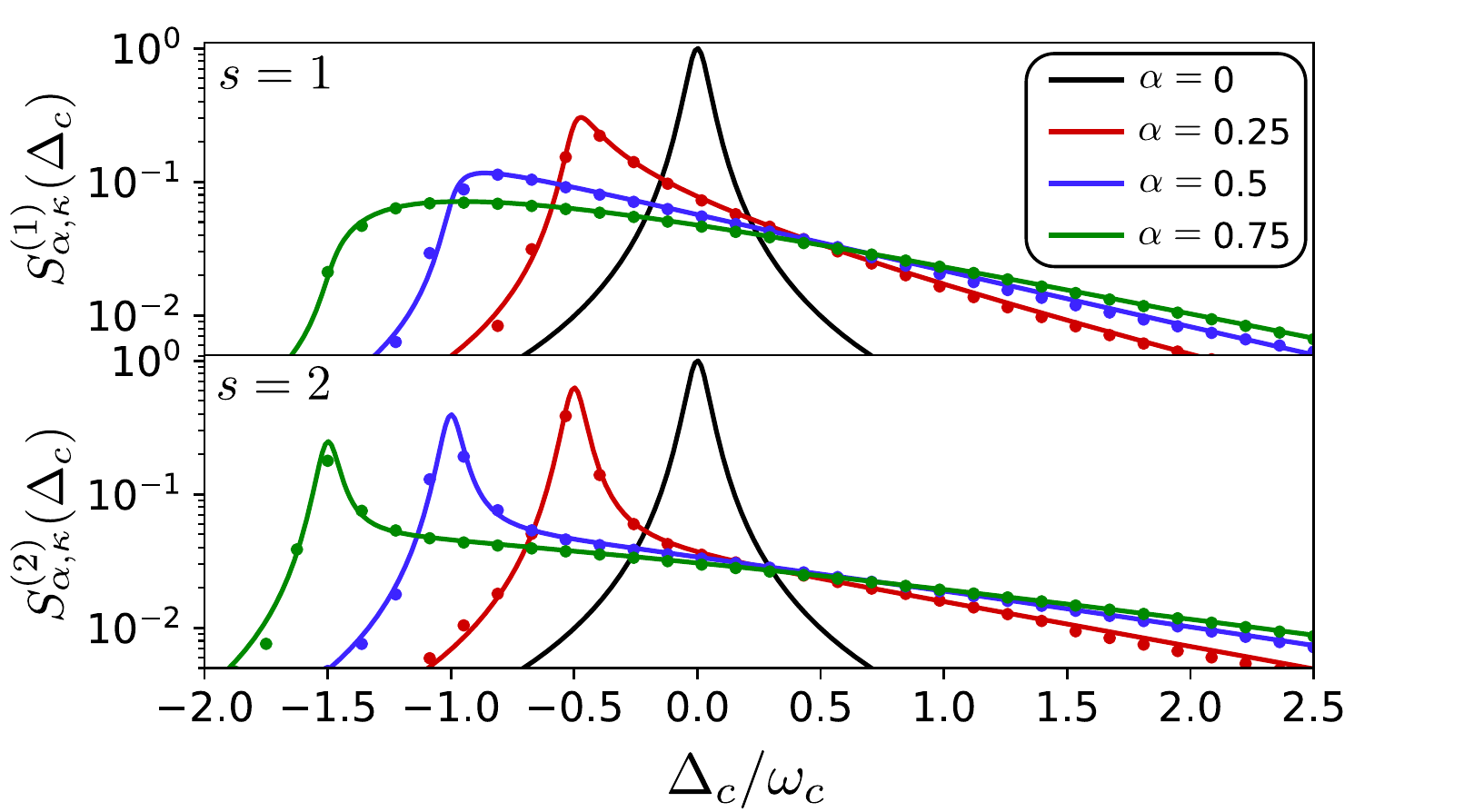}
    \caption{Excitation spectrum $S^{(s)}_{\alpha, \kappa}(\Delta_c)$ obtained in the weak-driving limit from Eq.~\eqref{eq:SFormula} (solid) and exact numerical results for finite driving, $\eta = 0.05\omega_c$, using TEMPO (dots). In both plots $\kappa = 0.1\omega_c$.}
    \label{fig:Fig2}
\end{figure}

Figure~\ref{fig:Fig2} shows this excitation spectrum for $s=1,2$ and different coupling strengths.
We find good agreement between the results of the perturbation theory with exact numerics using a path integral approach~\cite{Note1}.
We observe a shift and a reduction of the strength of the original cavity resonance and the emergence of a broad background due to the bath modes. 
There are, however, qualitative differences between the Ohmic and the super-Ohmic cases, which can be understood analytically for moderate coupling $\alpha<1$ and small Markovian losses, $\kappa\ll \omega_c$~\cite{Note1}.
In this regime, the spectrum in the super-Ohmic case ($s=2$) can be written as the \textit{sum} of two contributions
\begin{equation}\label{eq:S2}
S_{\alpha, \kappa}^{(2)}(\Delta_c) = e^{-2\alpha}\frac{(\kappa/2)^2}{\Delta^2 + (\kappa/2)^2} +\mathcal{S}_{B}(\Delta),
\end{equation}
as is familiar from the excitation spectra of two-level defects in solids~\cite{Huang50,Mahan65,Krummheuer02,WilsonRae2002,Roy11}. 
Although the cavity resonance is reduced and shifted to $\Delta_{\rm res}^{(s=2)}=-\Delta_p$,  there is---for all coupling parameters---a distinct quasi-photon peak, which sits on top of a broad background described by $\mathcal{S}_{B}(\Delta)$~\footnote{Note that through the relation $\mathcal{S}_{B}(\Delta)= e^{-2\alpha}\sum_{n=1}^{\infty}(2\alpha)^nS_{\alpha=n/2, \kappa}^{(1)}(\Delta)/n!$, the spectrum in the super-Ohmic case can be expressed in terms of that of the Ohmic spectrum given in Eq.~\eqref{eq:S1}. 
For small $\alpha$, we obtain $\mathcal{S}_{B}(\Delta)\simeq  2\alpha e^{-2\alpha} S_{\alpha=1/2, \kappa}^{(1)}(\Delta)$.}. 
In contrast, the spectrum for the Ohmic bath ($s=1$) is given by 
\begin{equation}\label{eq:S1}
S_{\alpha, \kappa}^{(1)}(\Delta_c)  \simeq 
 \left[\frac{(\kappa/2)^2}{\Delta^2+(\kappa/2)^2}\right]^{\delta} \times \mathcal{F}_B(\Delta),
\end{equation}
where $\delta=1/2-\alpha$ is the distance from the Toulouse point and $\mathcal{F}_B(\Delta)=\left(\frac{\kappa}{2\omega_c}\right)^{1-2\delta} \frac{\Gamma(2\delta)}{\text{e}^{\Delta/\omega_c}} \sin\left[\pi\delta+2\delta\arctan\left(\frac{2\Delta}{\kappa}\right)\right]$ \cite{Toulouse69,Guinea85}. 
Therefore, in the Ohmic case the cavity resonance is substantially modified by the interaction with the bath. 
For $\alpha < 1/2$, there is still a quasi-photon-like peak centered at $\Delta^{(s=1)}_{\rm res} \simeq -\Delta_p +\kappa\tan\left[\pi(1-2\delta)/(1+2\delta)\right]/2$,
but with a strongly asymmetric line-shape \cite{Mahan65,Vorrath05}. For $\alpha>1/2$ the quasi-photon peak vanishes completely and we are left with  purely polaronic excitations.

\emph{Photon Blockade.}---The excitation spectrum $S(\Delta_c)$ only captures the single-photon physics. 
To understand how the different baths can result in photon nonlinearities we now consider the two-photon correlation function $g^{(2)}(0)=\lim_{t\rightarrow\infty}\average{ c^{\dagger}c^{\dagger}cc}_t/\average{c^{\dagger}c}^2_t$. 
This quantity measures the relative weights of two-photon and single-photon excitations, satisfying $g^{(2)}(0)=1$ for a coherent state and $g^{(2)}(0)=0$ for a two-level system. 
The effect of anti-bunching or \emph{photon blockade}~\cite{Imamoglu97,birnbaum2005,Tian92} with $g^{(2)}(0)<1$ cannot be explained by a classical description of the field~\cite{MilburnWallsBook,ZollerGardiner04} and indicates a strong nonlinearity, which prevents a second photon from entering the cavity.

\begin{figure}
\includegraphics[width=\columnwidth]{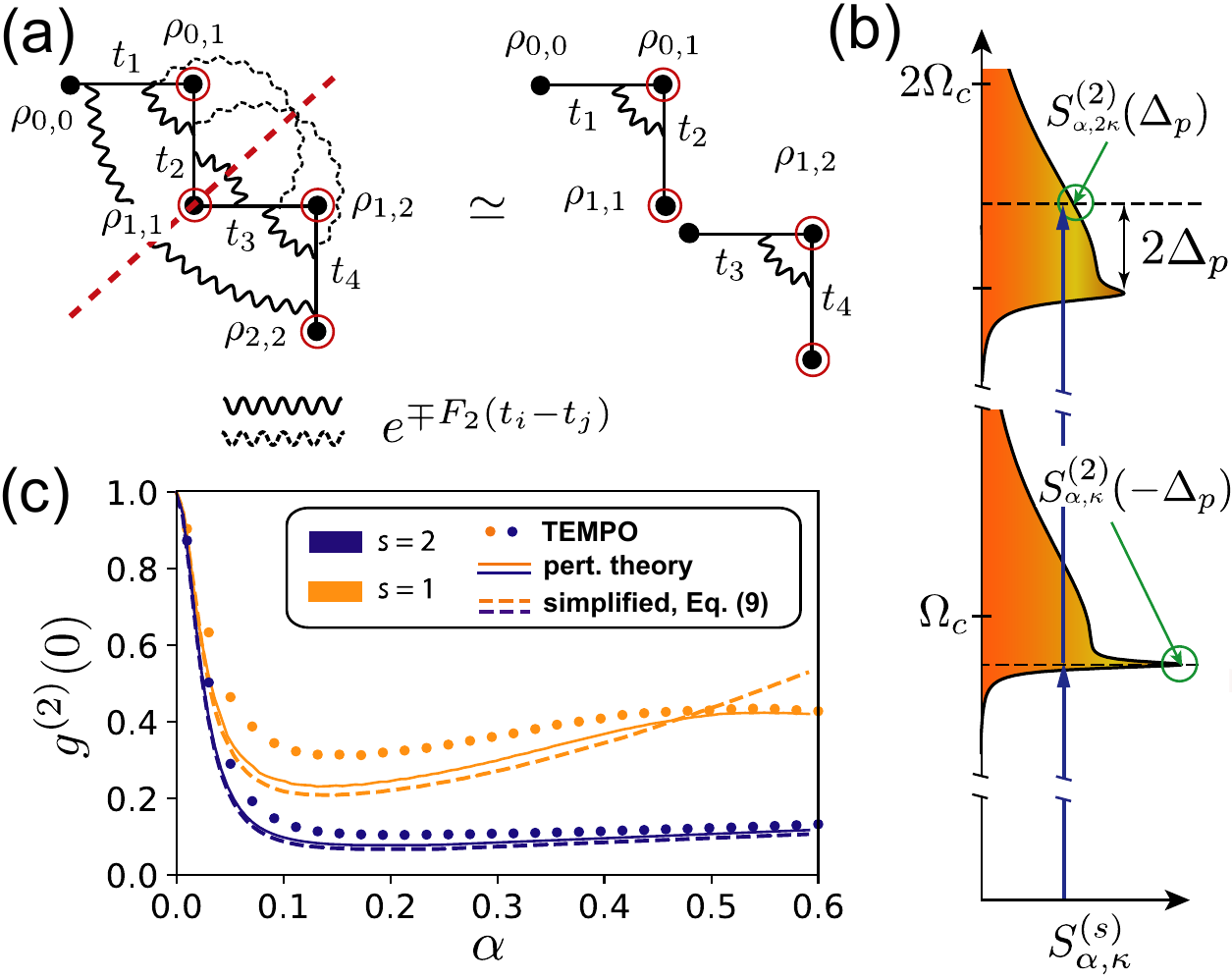} 
    \caption{(a) Diagrammatic representation~\cite{Thorwart05} of the four-point correlator $\langle e^{iP(t_1)}e^{iP(t_3)}e^{-iP(t_4)}\text{e}^{-iP(t_2)}\rangle_{\rm th}$, written as a product of six two-time functions $e^{\pm F_2(t_i-t_j)}$. 
The right panel shows the approximate form for resonant driving, where all correlations between the first and the second photon are neglected. (b) Pictorial representation of the simplified formula, Eq.~\eqref{eq:g20simple}, for two-photon absorption.
     (c) Plot of the  two-photon correlation $g^{(2)}(0)$ for $\Delta_c = \Delta_{\mathrm{res}}^{(s)}$, $\kappa = 0.1\omega_c$ and $\eta = 0.05\omega_c$.  } 
\label{fig:Fig3}
\end{figure}

Within perturbation theory we may use the diagrammatic representation of the possible processes shown in Fig.~\ref{fig:Fig1}(c) to calculate $g^{(2)}(0)$.
The reduced density matrix of interest, $\rho_{2,2}$, is connected to the unperturbed state $\rho_{0,0}$ through different paths involving four operators $\sim \eta \text{e}^{\pm i P}$.  
After averaging over the bath, the two-photon occupation $\average{c^\dag c^\dag c c}\simeq 2\tr_B\{\rho_{2,2}\}$ is thus given in terms of integrals over four-point polaron correlators of the form $\langle \prod_{i=1}^4 e^{i \sigma_i  P(t_i)} \rangle_{\rm th}=\prod_{i<j} e^{\sigma_i\sigma_j F_2(t_i-t_j)}$ ~\cite{Note1}, where $\sigma_i=\pm 1$ and $\sum_j \sigma_j=0$ (see also related calculations in Refs.~\cite{Egger94, Hu98, Wuerger98,Furusaki98, Thorwart05, Rabl2011}).
As illustrated in Fig.~\ref{fig:Fig3}(a), each of the six factors $e^{\pm F_2(t_i-t_j)}$ in this product affects one of the coherences involved in a given two-photon excitation path. As a result, the full expression for $g^{(2)}(0)$~\cite{Note1} is considerably more involved than the single-photon spectrum and in general not very enlightening. However, for the most relevant case of resonant driving, $\Delta_c=\Delta_{\rm res}^{(s)}$, and a good cavity, $\kappa\ll \omega_c$, reservoir-induced correlations between the first and second excitation can be neglected~\cite{Note1}, as depicted in Fig.~\eqref{fig:Fig3}(a).
Physically, this approximation means that the absorption of the second photon is independent of the first and so can be treated as a single photon event which is off-resonant by $2\Delta_p$ and with an increased decay rate $2\kappa$. 
This simplification allows us to express the two-photon correlations solely in terms of the single-photon spectra,
\begin{equation} \label{eq:g20simple}
g^{(2)}(0)\simeq \frac{S_{\alpha, 2\kappa}^{(s)}\left(\Delta_{\rm res}^{(s)}+2\Delta_p\right)}{S_{\alpha, \kappa}^{(s)}\left(\Delta_{\rm res}^{(s)}\right)}.
\end{equation}
Combined with Eqs.~\eqref{eq:S2} and~\eqref{eq:S1}, this result provides a closed analytic expression for $g^{(2)}(0)$, which is expected to hold as long as the quasi-photon picture is valid.

In Fig.~\ref{fig:Fig3}(a) we plot the full analytic result for the two-photon correlation function in the weak driving limit, together with the approximation in Eq.~\eqref{eq:g20simple} and a numerically exact simulation of $g^{(2)}(0)$ for finite $\eta$.
Naively, the continuum of bath states always guarantees a resonant state for a second photon, but all methods show photon blockade at all values of $\alpha$ for both types of environment. 
This observation can be explained from the simplified expression, Eq.~\eqref{eq:g20simple}. This states  that $g^{(2)}(0)<1$ can be obtained whenever the single-photon spectrum is monotonically decreasing above $\Delta_{\rm res}^{(s)}$ (as long as the $\kappa$ dependence can be ignored). 
This is clearly the case for the super-Ohmic bath, where a sharp quasi-photon resonance exists for all couplings. 
However, even for an Ohmic bath in the overdamped regime, $\alpha\gtrsim 1/2$, the existence of a shallow maximum at  $\Delta_{\rm res}^{(1)}$ can still lead to a significant nonlinear effect.
While for  $\alpha>1/2$ the result of Eq.~\eqref{eq:g20simple} is no longer accurate, the full analytic and numerical results confirm the validity of this picture beyond this point.

\emph{Photonic Rabi Oscillations.}---Our finding of $g^{(2)}(0)<1$ in the weakly driven regime indicates an environment-induced mechanism that suppresses excitations of higher photon number states. 
It is thus intriguing to ask, whether this mechanism can isolate a photonic two-level subspace and drive Rabi oscillations between the (dressed) photon states $\ket{0}$ and $\ket{1}$. 
To answer this question it is necessary to consider a strongly driven system, $\eta\gg \kappa$, where $n_0\gg1$ and the perturbative approach above is no longer applicable. 
Therefore, we perform numerical simulations, using the recently developed time-evolving matrix product operator (TEMPO) algorithm~\cite{Strathearn2018}. 
This algorithm treats the dynamics of the bath exactly and uses a matrix-product operator ansatz with finite bond dimension to simulate the resulting time-nonlocal dynamics for the cavity mode  efficiently, allowing us to keep up to eight photons in the cavity Hilbert space to ensure convergence. In Figs.~\ref{fig:Fig2} and~\ref{fig:Fig3} we applied this technique to compare the weak-driving results for $S^{(s)}_{\alpha,\kappa}(\Delta_c)$ and $g^{(2)}(0)$ with a full simulation at finite $\eta$.

\begin{figure}
    \includegraphics[width=\columnwidth]{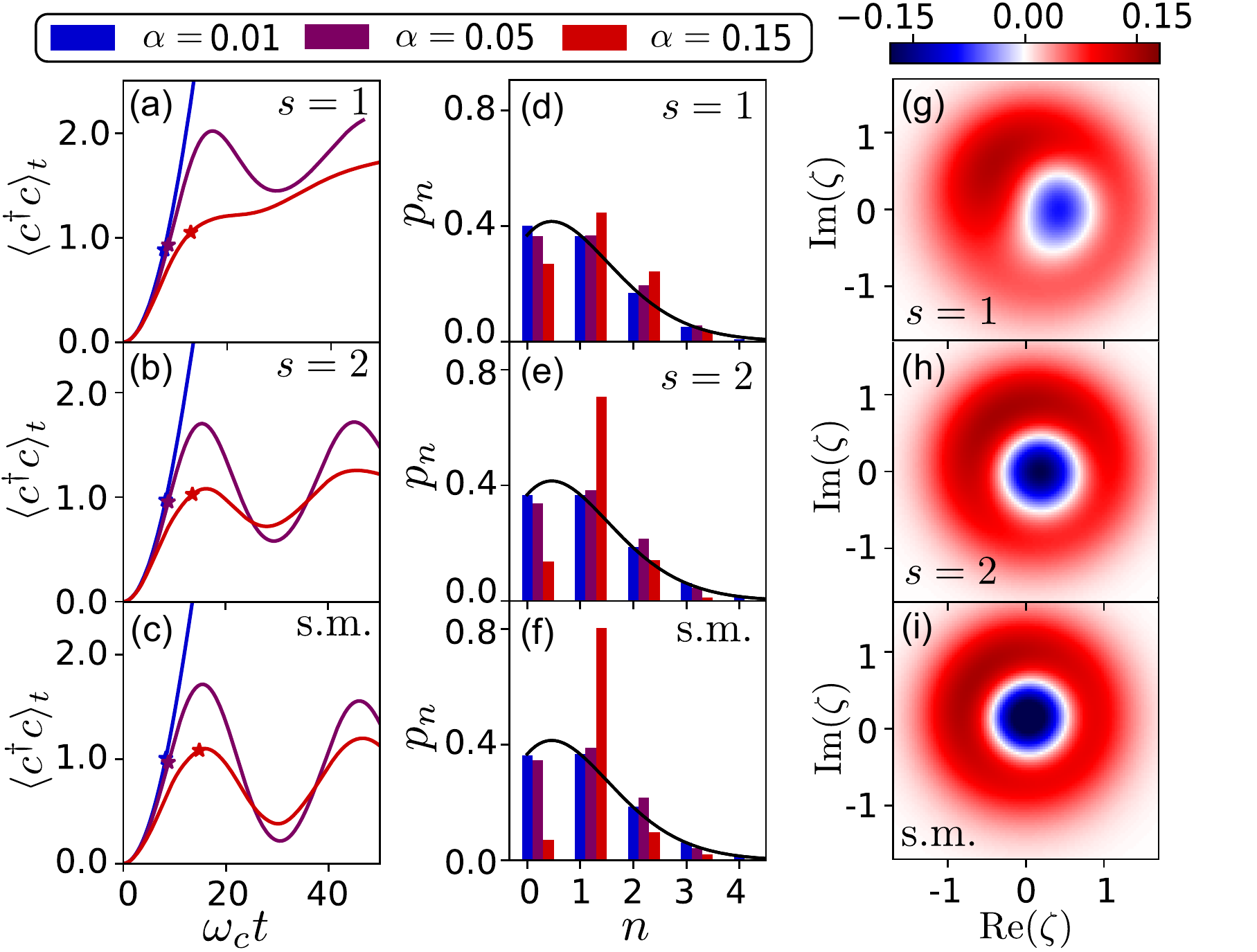} 
    \caption{Numerical simulation of Eq.~\eqref{eq:ME} with $\kappa/\omega_c = 0.01$, $\eta/\kappa = 25$ and different coupling parameters. The Ohmic ($s=1$) and super-Ohmic ($s=2$) environments are compared with that of a single bath mode (s.m.) of frequency $\omega_c$. (a)-(c) Evolution of the mean photon number of a resonantly driven cavity, $\Delta_c=\Delta_{\rm res}^{(s)}$.  (d)-(f) Photon number distribution, $p_n$, evaluated at times $t_*$ [marked by stars in panels (a)-(c)] for which $p_1$ is maximized. (g)-(i) Plot of the Wigner function, $W(\zeta,\zeta^*)$, evaluated at times $t_*$ and for $\alpha=0.15$. 
    }
    \label{fig:Fig4}
\end{figure}

Figure~\ref{fig:Fig4} summarizes the numerical results for the transient dynamics of a strongly driven system, a regime where other methods are not available. 
The plots compare the evolution of the cavity population for Ohmic and super-Ohmic baths to that where the bath is replaced by a single mode of frequency $\omega_c$ and coupling $g$. 
In all cases the driving field is tuned to the single-photon resonance, $\Delta_c=\Delta_{\rm res}^{(s)}$, and the value of the coupling for the single mode is chosen to match the nonlinearity of the other cases so that $g^2/\omega_c^2 = 2\alpha$.
For very small couplings the cavity photon number simply increases monotonically, $\average{c^\dag c}_t\simeq  n_0e^{-\kappa t}(1-e^{\kappa t/2})^2$, as expected for a resonantly driven oscillator. 
We may only simulate this for short times before the dynamics reaches the highest level in the truncated Hilbert space.
For slightly larger coupling parameters the dynamics changes drastically and the occupation number starts to oscillate at low photon numbers, $\average{c^\dag c}_t \sim O(1)\ll n_0=625$.
Consistent with the previous discussion, these oscillations have higher visibility in the super-Ohmic case, where the effect of the continuous bath is similar to that of a single mode.
With increasing coupling strength the bath-induced decoherence becomes more important and the oscillations are rapidly damped. 
As a result, there exists an optimal range of coupling parameters between the linear and the damped regime where the oscillations are most pronounced. 

To clarify the nature of these oscillations, we take snapshots of the photon number distribution $p_n=\tr_B\bra{n}\rho\ket{n}$ at times when $p_1$ is maximized. Even in the Ohmic case at $\alpha=0.15$, we observe significant deviations from a Poisson distribution.
In the super-Ohmic case the distribution shows a dominant peak at $p_1$.
We further  plot the Wigner function $W(\zeta,\zeta^*)$  for these states
\cite{MilburnWallsBook,Agarwalbook12}, which are close to that of a single-photon Fock state, exhibiting a large negative region which demonstrates the nonclassical nature of the state.
This analysis shows that for appropriate parameters the dissipative cavity dynamics shows Rabi-like oscillations between the states $\ket{0}$ and $\ket{1}$, where for the investigated range of $\alpha$ the Rabi-frequency $\Omega_R\approx \eta$ is close to the bare driving strength. 

\emph{Conclusion.}---
In summary, we have studied the dynamics and steady states of a driven oscillator under the influence of a low-frequency bosonic bath. Specifically, we have addressed the fundamental competition between emergent nonlinearities and decoherence in this model and we have shown that strong coupling to an unstructured environment can turn a harmonic  oscillator into an effective two-level system undergoing Rabi-like oscillations. These results will be important, for example, for engineering novel types of optical nonlinearities through the coupling to a broad-band continuum of mechanical modes.

\textit{Acknowledgements.}---We thank A.~Strathearn, J.~J.~Garcia-Ripoll, L.~Henriet, A. Sipahigil and O. Painter for stimulating discussions. 
This work was supported by the Austrian Science Fund (FWF) through the START Grant No.~Y 591-N16 and the DK CoQuS, Grant No.~W 1210. PK acknowledges support from an ESQ fellowship of the Austrian Academy of Sciences (\"OAW).

%

\widetext
\clearpage
\begin{center}
    \textbf{\large Supplementary material for:  Environment-Induced Rabi Oscillations in the Optomechanical Boson-Boson Model}
\end{center}
\setcounter{equation}{0}
\setcounter{figure}{0}
\setcounter{table}{0}
\setcounter{page}{1}
\makeatletter
\renewcommand{\theequation}{S\arabic{equation}}
\renewcommand{\thefigure}{S\arabic{figure}}
\renewcommand{\bibnumfmt}[1]{[S#1]}
\renewcommand{\citenumfont}[1]{S#1}

\section{Derivation of  the Master Equation}

In this section we will give a detailed derivation of the master equation used in the main text. 
The full model describes the single cavity mode coupled to two bosonic baths, one for the structured, non-Markovian environment and another which describes the weak linear interaction with a broad-band electromagnetic vacuum. 
The total Hamiltonian is then given by
\begin{equation}
H_{\text{full}} = {\Omega}_c c^{\dagger}c + g c^{\dagger}c X + H_{B}  + \underbrace{\int d\omega \, G(\omega)  (c^{\dagger} c_\omega + c_\omega^{\dagger} c)}_{=H_{c\text{-}\bar{B}}} + \underbrace{\int d\omega \, \omega\, c_\omega^{\dagger}c_\omega}_{=H_{
        \bar{B}}}.
\end{equation}
The interaction of the cavity with the electromagnetic environment is modeled by a bath $\bar{B}$ of bosonic modes with operators $c_\omega$, which obey $[c_\omega,c^\dag_{\omega'}]=\delta(\omega-\omega')$.

All other symbols have the same meaning as in the main text. Note that by writing $H_{c-\bar B}$ we have already made a rotating wave approximation and omitted terms $\sim c^\dag c^\dag_\omega$ and $\sim c c_\omega$. This is justified, since the coupling to bath $\bar B$ is assumed to be sufficiently weak such that only energy-conserving processes are relevant.  In the following we further assume that the temperature $T$ is low enough to neglect the thermal occupation of the near-resonant bath modes, $n_{\rm th}(\Omega_c)=1/(e^{\hbar \Omega_c/(k_BT)}-1)\ll 1$. Both assumptions are well justified for optical cavity modes, but also for superconducting microwave resonators operated at dilution refrigerator temperatures. 

In this setting a classical field driving the cavity at frequency $\omega_d$ is modeled by an initial bath state where the mode at the driving frequency $\omega=\omega_d$ is the displaced vacuum state $\rho_{\bar{B}}(0) = \ket{ \beta}\bra{\beta}$ with $\ket{ \beta} =  \mathcal{D}_{c_{\omega_d}}(\beta)\ket{0}$. 
Here, $\mathcal{D}_a(\beta)=\exp(a^{\dagger}\beta - \beta^* a)$ is the usual displacement operator. 
This initial displacement can be eliminated by a unitary transformation $U_\beta=e^{+iH_{\bar{B}}t}\mathcal{D}_{c_{\omega_d}}(\beta)e^{-iH_{\bar{B}}t}$ such that we obtain
\begin{equation}
\tilde H_{\text{full}}(t) = {\Omega}_c c^{\dagger}c + \frac{\eta}{2}(ce^{i\omega_d t} + c^{\dagger}e^{-i\omega_d t})+ g c^{\dagger}c X + H_{B} + H_{c\text{-}\bar{B}} + H_{\bar{B}},
\end{equation}
where we defined $\eta = 2G(\omega_d)\beta$, and the bath is now initialized in the vacuum state. The explicit time dependence can be eliminated by going to a rotating frame $H_{\rm rf} = U(t)\tilde{H}(t)U^{\dagger}(t)-i \dot U(t) U^\dag(t)$, where $U(t)=\mathrm{exp}\left[-i\omega_d t (c^{\dagger}c+\int d\omega \,  c_\omega^{\dagger}c_\omega)\right]$. In this frame the Hamiltonian is of the form of the boson-boson model introduced in the main text,
\begin{align}
H_{\text{rf}} &= -\Delta_c c^{\dagger}c+\frac{\eta}{2}(c + c^{\dagger})+g  c^{\dagger}c X + H_{B} + \int d\omega \, G(\omega)  (c^{\dagger} c_\omega + c_\omega^{\dagger} c)+ \int d\omega \, (\omega-\omega_d) \, c_\omega^{\dagger}c_\omega 
\\
& = H_{\rm BBM} + \sqrt{\frac{\kappa}{2\pi}}\left(c^{\dagger}C + C^{\dagger}c\right) + \int d\omega \, (\omega-\omega_d) \, c_\omega^{\dagger}c_\omega,   
\end{align}
where we defined the detuning of the drive from the cavity, $\Delta_c = \omega_d-{\Omega}_c$. By going from the first to the second line we have assumed that the coupling $G(\omega)$ varies smoothly as a function of frequency and introduced the decay rate
\begin{equation}
\kappa = 2\pi |G(\omega=\Omega_c)|^2
\end{equation}
and the collective bath operator $C=\int d\omega \, c_\omega$.
By then performing the standard Born-Markov approximation~\cite{suppBreuer_book} on this bath we can eliminate the modes of the electromagnetic environment to arrive at the master equation for the reduced dynamics of the cavity and non-Markovian bath,
\begin{equation}\label{eq:MESupp}
\dot\rho  = -i[H_{\rm BBM}, \rho ] + \frac{\kappa}{2}(2c\rho c^\dag -c^\dag c \rho - \rho c^\dag c),
\end{equation}
which is the same as that used throughout the main text. 

Note that for the validity of the whole derivation it is enough that all assumptions that we have made apply for the bath modes within a band $\omega\in [\Omega_c-\Delta_{\bar B},\Omega_c+\Delta_{\rm \bar B}]$ of width $\Delta_{\bar B}$ around the resonance frequency.  The validity of the Born-Markov approximation then requires that this bandwidth is large compared to all the other frequency scales, $\Delta_{\bar B}\gg \kappa, \Delta_c, \omega_c,\eta$. This is typically the case for the quantum optical and circuit QED settings considered in this work.

\section{Perturbation Theory}

In this section we give details of the perturbative treatment of the boson-boson model, which is valid in the weak driving limit, $\eta/\kappa\rightarrow 0$. This approach is used in the main paper to evaluate the steady-state photon number and two-photon correlations.
The starting point is the Hamiltonian in the polaron frame, as in the main text 
\begin{equation}
H_P  = -\Delta_c c^{\dagger}c -\Delta_p c^{\dagger}c\,c^{\dagger}c +\frac{\eta}{2}(e^{iP}c+c^{\dagger}\text{e}^{-iP})+H_B,
\end{equation}
where $P = ig\sum_k (\xi_k^*b^\dag_k-\xi_kb_k)/\omega_k$ is the collective momentum operator. Transforming to the polaron frame also adjusts the jump term in the master equation
\begin{equation}
\dot{\rho}= -i[H_P,\rho ] + \frac{\kappa}{2}\left(2c e^{iP}\rho e^{-iP} c^\dag -c^\dag c \rho - \rho c^\dag c\right).
\end{equation}
In the following we will consider the limit of weak driving, $\eta\ll \kappa$, such that we perturb weakly around the exactly solvable limit where $\eta=0$.

To go ahead with this perturbation theory we begin by transforming to the interaction picture with respect to the bath Hamiltonian $\rho(t) \rightarrow e^{-iH_Bt}\rho(t)e^{iH_Bt}$ giving the master equation
\begin{equation}
\dot{\rho} =-i[H_P(t),\rho ] +\frac{\kappa}{2}\left(2c e^{iP(t)}\rho e^{-iP(t)} c^\dag -c^\dag c \rho - \rho c^\dag c\right),
\end{equation}
where $H_P(t)=e^{iH_Bt}(H_P-H_B)e^{-iH_Bt}$ and $P(t) = e^{iH_Bt}Pe^{-iH_Bt}$.  From this equation we can derive an infinite hierarchy of equations for the reduced bath density operators $\rho_{m,n}=\langle m \vert \rho \vert n\rangle$,
\begin{equation}\label{suppeq:dt_rhomn}
\begin{split}
\dot{\rho}_{m,n}(t) & \simeq \left(i\Delta_{mn} -\frac{\kappa_{mn}}{2} \right)\rho_{m,n}(t)+ \kappa \sqrt{m+1}\sqrt{n+1} e^{iP(t)}  \rho_{m+1,n+1}(t)e^{-iP(t)}\\
&-i\frac{\eta_{m+1}}{2}e^{iP(t)}  \rho_{m+1,n}(t) -i\frac{\eta_m}{2}e^{-iP(t)}  \rho_{m-1,n}(t)+i\frac{\eta_n}{2} \rho_{m,n-1}(t)e^{iP(t)} + i\frac{\eta_{n+1}}{2}\rho_{m,n+1}(t) e^{-iP(t)},
\end{split}
\end{equation}
where $\eta_n=\eta \sqrt{n}$ and $\Delta_{mn}$ and $\kappa_{mn}$ are defined in the main text. 
In the case where the cavity is initially in the vacuum state we have $\rho_{0,0}(0)=\rho_{\rm th}$ and $\rho_{n,m}(0)=0$ otherwise. This means that by iteratively solving the coupled equations for small but finite $\eta$ we obtain the scaling
\begin{equation}
\rho_{m,n}(t) \sim O\left( \eta^{m+n}\right).
\end{equation}
In particular, this scaling implies that on the right hand side of Eq.~\eqref{suppeq:dt_rhomn} we only need to keep terms $\sim \rho_{m,n}$, $\sim \rho_{m-1,n}$ and  $\sim \rho_{m,n-1}$. Neglecting all other terms gives Eq.~\eqref{eq:rhonm} in the main text.
The density matrix $\rho_{m,n}$ can then be related to those with smaller $n,m$ by formally integrating this equation,
\begin{equation}\label{suppeq:RecursionRelation}
\rho_{m,n}(t)\simeq \int_0^td\tau\,\mathcal{G}_{m,n}(t-\tau)
\left[ 
i\frac{\eta_{n}}{2}\rho_{m,n-1}(\tau)e^{iP(\tau)} -i\frac{\eta_{m}}{2}e^{-iP(\tau)}\rho_{m-1,n}(\tau)
\right],
\end{equation}
where we have defined 
\begin{equation}
\mathcal{G}_{mn}(t)=e^{(i\Delta_{mn}-\kappa_{mn}/2)t}.
\end{equation}
By recursive iteration we obtain an expression for $\rho_{m,n}(t)$ depending only on $\rho_{0,0}(t)$. For weak driving, $n_0\ll 1$, the initial state is barely altered and so $\rho_{0,0}(t)\simeq \rho_{0,0}(0)$. 

\subsection{Steady-State Photon Number}
We first employ the perturbation theory developed above to compute the steady-state cavity photon number
\begin{equation}
S^{(s)}_{\alpha,\kappa}(\Delta_c)\equiv \underset{t\rightarrow \infty}{\lim}\langle c^{\dagger}c\rangle_{\mathrm{t}}/n_0 \simeq {\rm tr}_B\{\rho_{1,1}(t\rightarrow \infty)\}/n_0 +O(n_0^{1/2}).
\end{equation}
By iterating Eq.~\eqref{suppeq:RecursionRelation} twice we readily obtain
\begin{equation}
\begin{split}
p_1(t\rightarrow \infty) 
& \simeq 
\frac{\eta^2}{4}\int_0^{t\rightarrow \infty}dt_2\int_0^{t_2}dt_1\, \left[ \mathcal{G}_{11}(t-t_2) \mathcal{G}_{01}(t_2-t_1)\langle e^{iP(t_1)}e^{-iP(t_2)}\rangle  + \mathrm{c.c.} \right] \\
&=\frac{\eta^2}{2\kappa}\mathrm{Re}\int_0^{\infty}d\tau\, e^{(i\Delta+\kappa/2)\tau}\langle e^{iP(\tau)}e^{-iP(0)}\rangle,
\end{split}
\end{equation}
where $\Delta=\Delta_c+\Delta_p$. After normalization with respect to the empty cavity photon number, $n_0$, we obtain the result from the main text
\begin{equation}
S^{(s)}_{\alpha,\kappa}(\Delta_c) = \frac{\kappa}{2}\Re\int_0^{\infty}d\tau \, e^{(i\Delta+\kappa/2)\tau} e^{-F_2(\tau)}.
\end{equation}
Here we have used that for thermal (i.e. Gaussian) states 
\begin{equation} \label{eq:F2}
\begin{split}
\langle e^{iP(\tau)}e^{-iP(0)}\rangle_{\rm th} & = \langle e^{i(P(\tau)-P(0))}\rangle_{\rm th} e^{\frac{1}{2}[P(\tau),P(0)]}=e^{-\frac{1}{2}\langle(P(\tau)-P(0))^2\rangle_{\rm th} } e^{\frac{1}{2}\langle[P(\tau),P(0)]\rangle} \\
& = e^{-\langle P^2\rangle_{\mathrm{th}}+ \frac{1}{2}\langle \left\{ P(\tau),P(0) \right\}\rangle_{\mathrm{th}} + \frac{1}{2}\langle [ P(\tau),P(0)) ]\rangle_{\mathrm
        th}} =e^{-(\langle P^2\rangle_{\rm th} - \langle P(\tau)P(0)\rangle_{\rm th})} = e^{-F_2(\tau)},
\end{split}
\end{equation}
from which we obtain the standard result 
\begin{equation}
\begin{split}
F_2(\tau)= &g^2  \sum_k \frac{|\xi_k|^2}{\omega_k^2}  \left[ (2n(\omega_k)+1)(1-\cos(\omega_k \tau))   + i \sin(\omega_k \tau)  \right]\\
=& \int_0^\infty d \omega \, \frac{J(\omega)}{\omega^2} \left[ \coth\left(\frac{\hbar \omega}{2k_BT} \right) (1-\cos(\omega \tau))+  i \sin(\omega \tau)   \right].
\end{split}
\end{equation}
\subsubsection*{Zero-temperature limit}
In the following we will focus on the limit $T\rightarrow 0$ such that we obtain \cite{suppLeggett1987,suppweiss_book}
\begin{equation} \label{suppeq:F2corr}
F_{2}(t) = 2\alpha \Gamma(s-1)\left(1-(1+i\omega_c t)^{1-s}\right)=
\begin{cases}
2\alpha \ln(1+i\omega_c t) = \alpha\, \mathrm{ln}(1+\omega_c^2 t^2) + i 2\alpha \mathrm{arctan}(\omega_c t)  & s = 1,\\
2\alpha \left(1-\frac{1}{1+i\omega_c t}\right)=2\alpha\left( 1-\frac{1}{1+(\omega_ct)^2}\right) + i  \frac{2\alpha\omega_c t}{1+(\omega_c t)^2} & s= 2.
\end{cases} 
\end{equation}
In the Ohmic limit, $s=1$, we then obtain an analytical solution for the above integral 
\begin{equation}
\begin{split} \label{suppeq:ST0alpha}
S^{(s=1)}_{\alpha,\kappa}(\Delta_c) & = \frac{\kappa}{2}\Re\int_0^{\infty}d\tau\, \frac{e^{(i\Delta-\kappa/2)\tau}}{(1+i\omega_c \tau)^{2\alpha}} \\
& = \frac{\kappa}{2}\Re\left[  (i\omega_c)^{-2\alpha}\left( \frac{\kappa}{2}-i\Delta \right)^{2\alpha-1}e^{-(\Delta+i\kappa/2)/\omega_c} \Gamma(1-2\alpha,-(\Delta+i\kappa/2)/\omega_c)\right],
\end{split}
\end{equation}
where we used the result that $\int_0^{\infty}d\tau\,(\beta+\tau)^{\nu}e^{-\mu \tau} = \mu^{-\nu-1}e^{\beta\mu}\Gamma(\nu+1,\beta\mu)$, which requires $\Re(\mu)>0$ and $\vert\mathrm{arg}(\beta)\vert< \pi$. In this expression $\Gamma(\alpha,x)\equiv \int_x^{\infty}d\tau\,e^{-\tau}\tau^{\alpha-1}$ is the incomplete Gamma function \cite{suppGradshteyn07}.

In the limit of a good cavity and close to resonance, $\vert\Delta\vert,\kappa \ll \omega_c $, we can approximate $\Gamma(1-2\alpha,-(\Delta-i\kappa/2)/\omega_c)\simeq \Gamma(1-2\alpha)$ and find
\begin{equation}
\begin{split}
S^{(1)}_{\alpha,\kappa}(\Delta_c) & \simeq \frac{\kappa}{2}\frac{\Gamma(1-2\alpha)}{\omega_c^{2\alpha}}e^{-\Delta/\omega_c}\left[\Bigl( \frac{\kappa^2}{4}+\Delta^2 \Bigr)^{\alpha-\frac{1}{2}}\mathrm{cos}\left( \pi\alpha -(1-2\alpha)\mathrm{arctan}\left(\frac{2\Delta}{\kappa} \right)\right)\right],
\end{split}
\end{equation}
in accordance with Eq.~\eqref{eq:S1} from the main body of the paper. Note that this approximation is still very accurate for larger values of $|\Delta|\sim \omega_c$, as long as the decay rate $\kappa$ remains small.  

The location of the quasi-photon like resonance for $\alpha \ll \frac{1}{2}$ is obtained by solving $\frac{d}{d\Delta_c}S^{(1)}_{\alpha,\kappa})(\Delta_c)=0$, which gives 
\begin{equation}
\mathrm{tan}\left[ \pi \alpha -(1-2\alpha)\mathrm{arctan}(2\Delta/\kappa)  \right] = \frac{\Delta^2+\frac{\kappa^2}{4}}{\frac{\kappa}{2}\omega_c(1-2\alpha)} + \frac{2\Delta}{\kappa} \simeq \frac{2\Delta}{\kappa}
\end{equation}
in the good cavity limit $\kappa \ll \omega_c$. This leads to 
\begin{equation}
\Delta^{(s=1)}_{\rm res} = -\Delta_p +\frac{\kappa}{2}\mathrm{tan}\left(\frac{\pi\alpha}{2(1-\alpha)}\right),
\end{equation} 
and when setting $\alpha=\frac{1}{2}-\delta$ we obtain the expression for the quasi-photon resonance mentioned in the main body of the paper. 

We now turn to the case of a super-Ohmic reservoir with $s=2$. 
The same procedure can be applied to obtain the results for any $s>1$ but for simplicity we will focus on this special case. 
The key observation is that the displacement correlation function can be expanded in terms of the series
\begin{equation}
e^{-F_2(\tau)} = e^{-2\alpha}\sum_{n=0}^{\infty}\frac{(2\alpha)^n}{n!}\frac{1}{(1+i\omega_c\tau)^{ n}}.
\end{equation}
In this case the steady-state photon number can be obtained by integration over each term in the series
\begin{equation}
S_{\alpha,\kappa}^{(s=2)}(\Delta_c) =e^{-2\alpha}\sum_{n=0}^{\infty}\frac{(2\alpha)^n}{n!}\frac{\kappa}{2}\Re\int_0^{\infty}d\tau\,\frac{e^{(i\Delta-\kappa/2)\tau}}{(1+i\omega_c\tau)^{ n}} =  e^{-2\alpha}\sum_{n=0}^{\infty}\frac{(2\alpha)^n}{n!}S^{(1)}_{n/2,\kappa}(\Delta_c),
\end{equation}
which can be used to express the whole series in terms of the $s=1$ steady-state photon number. The two first terms of the sum are
\begin{equation}
S_{0,\kappa}^{(1)}(\Delta)  =  \frac{\left(\frac{\kappa}{2}\right)^2}{\Delta^2+\left(\frac{\kappa}{2}\right)^2}, \;\;\;\;
S_{1/2,\kappa}^{(1)}(\Delta)  \simeq  \frac{\kappa}{2}\frac{e^{-\Delta/\omega_c}}{\omega_c^{2\alpha}}\left[ \frac{\pi}{2}+\mathrm{arctan}\left(\frac{2\Delta}{\kappa}\right)\right].
\end{equation}
The feature with the most physical implications, which can be clearly seen in the above expansion, is the existence and protection of the quasi-photon peak $S_{0,\kappa}^{(1)}(\Delta)$. This peak exists independently of the strength of the interaction with the bath, as opposed to the Ohmic case, where at strong coupling this well-defined cavity excitation vanishes.

\subsection{Two-Photon Correlation Function}

\begin{figure}
    \includegraphics[width=0.8\textwidth]{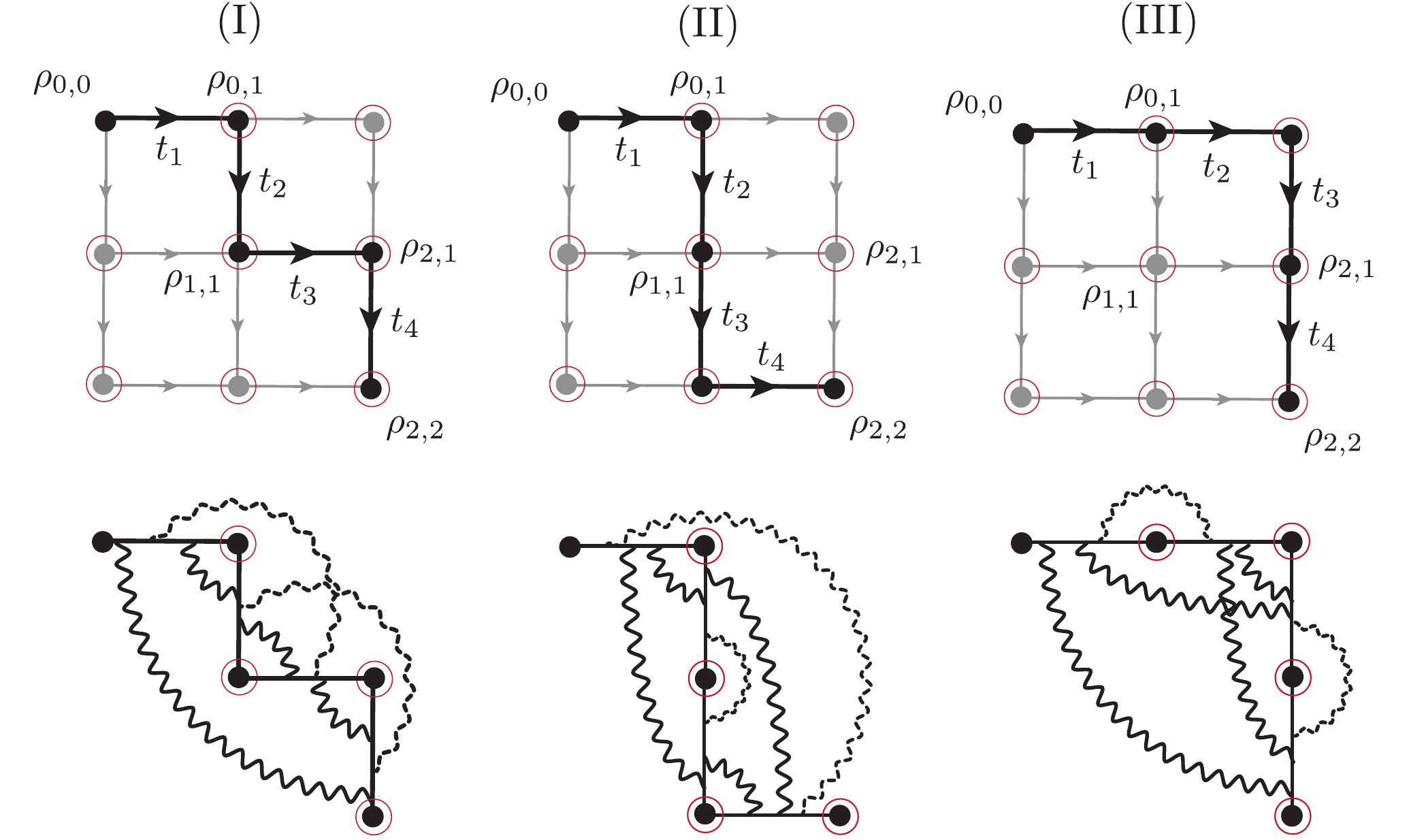} 
    \caption{Illustration of the three different excitation paths (their complex conjugate counterparts are not shown), which contribute to the full expression of $p_2$ in Eq.~\eqref{suppeq:p2full}. The second row indicates the from of the corresponding four-point polaron correlator when decomposed according to Eq.~\eqref{suppeq:Wick} 
        as $e^{-F_{4}(\{ t_j\})} = \prod_{i<j}^{4}e^{\sigma_i\sigma_j F_2(t_i-t_j)}$. The solid lines indicate factors of $e^{-F_2(t_i-t_j)}$ and the dashed lines factors of $e^{+F_2(t_i-t_j)}$.}
    \label{suppfig:suppFig1}
\end{figure}

In this section we derive the full result for the two-photon correlation function in the weak driving limit,
\begin{equation}
g^{(2)}(0)= \lim_{t\rightarrow \infty}\frac{\average{c^{\dagger}c^{\dagger}cc}}{\average{c^{\dagger}c}^2} \simeq \frac{2 {\rm tr}_B\{\rho_{2,2}(t\rightarrow \infty)\}}{\left[{\rm tr}_B\{ \rho_{1,1}(t\rightarrow \infty)\}\right]^2} + O( n_0^{1/2}).
\end{equation}
As shown in the diagrams in Fig.~\ref{suppfig:suppFig1}, there are three different paths (and their complex conjugates) to go from $\rho_{0,0}$ to $\rho_{2,2}$. By summing over all these possibilities we obtain for the two-photon population $p_2(t)={\rm tr}_B\{\rho_{2,2}(t)\}$
\begin{equation}\label{suppeq:p2full}
\begin{split}
p_2(t\rightarrow \infty) 
\simeq  \frac{\eta^4}{8} &\int_0^{t\rightarrow \infty}\mathrm{dt_4}\int_0^{t_4}dt_3\int_0^{t_3}dt_2\int_0^{t_2}dt_1 \\
&\Bigl[  \mathcal{G}_{22}(t-t_4)\mathcal{G}_{12}(t_4-t_3)\mathcal{G}_{11}(t_3-t_2)\mathcal{G}_{01}(t_2-t_1)\average{ e^{iP(t_1)}e^{iP(t_3)} e^{-iP(t_4)}e^{-iP(t_2)}}_{\rm th}  \qquad \,\,\,\,\,({\rm I}) \\
&  + \mathcal{G}_{22}(t-t_4)\mathcal{G}_{21}(t_4-t_3)\mathcal{G}_{11}(t_3-t_2)\mathcal{G}_{01}(t_2-t_1)\average{ e^{iP(t_1)}e^{iP(t_4)}e^{-iP(t_3)}e^{-iP(t_2)}}_{\rm th} \qquad ({\rm II}) \\
& + \mathcal{G}_{22}(t-t_4)\mathcal{G}_{12}(t_4-t_3)\mathcal{G}_{02}(t_3-t_2)\mathcal{G}_{01}(t_2-t_1)\average{ e^{iP(t_1)}e^{iP(t_2)}e^{-iP(t_4)}e^{-iP(t_3)} }_{\rm th} \qquad ({\rm III}) \\ & +\mathrm{c.c.}\Bigr]. 
\end{split}
\end{equation}
In order to evaluate the four-point correlation functions we generalize the expression in  Eq.~\eqref{eq:F2} and obtain a variant of Wick's theorem for displacement operators to express $2n$-point polaron correlation functions solely in terms of the function $F_2(\tau)$, which already appears in the evaluation of the  two-point correlation functions. We first derive the general relation
\begin{equation}
e^{-F_{2n}(\left\{ t_j\right\})} : =\average{\prod_{j=1}^{2n}e^{\sigma_j i P(t_j)}}_{\rm th} = e^{-(n\average{P^2}_{\rm th}+\sum_{i<j} \sigma_i \sigma_j \average{P(t_i)P(t_j)}_{\rm th})},
\end{equation}
where the exponent $F_{2n}(\left\{ t_j\right\})$ can be strongly simplified if $\sum_j \sigma_j=0$. In this case we obtain
\begin{equation}
\begin{split}
F_{2n}(\{ t_j\}) & = (n-n^2+n^2)\langle P^2\rangle_{\rm th} +\underbrace{\sum_{i<j}\sigma_i \sigma_j\langle P(t_i)P(t_j)\rangle_{\rm th}}_{n(n-1)\;\; \rm terms: \;\;\sigma_i\sigma_j=1} - \underbrace{\sum_{i<j}(-\sigma_i\sigma_j) \langle P(t_i) P(t_j)\rangle_{\rm th}}_{n^2\;\;\rm terms: \;\; \sigma_i \sigma_j=-1} \\
& =-\left( (n^2-n)\langle P^2\rangle_{\rm th}- \sum_{i<j}\sigma_i \sigma_j\langle P(t_i)P(t_j)\rangle_{\rm th}\right) + \left( n^2\langle P^2\rangle_{\rm th}-\sum_{i<j}(-\sigma_i\sigma_j) \langle P(t_i) P(t_j)\rangle_{\rm th} \right) \\
& = -\sum_{i<j} \sigma_i \sigma_j F_2(t_i-t_j).
\end{split}
\end{equation}
Therefore, the special combinations of displacement operators we encounter in Eq.~\eqref{suppeq:p2full} can be written in terms of factors of $e^{\pm F_2(t_i-t_j)}$ and in general we obtain
\begin{equation}\label{suppeq:Wick}
e^{-F_{2n}(\{ t_j\})} = \prod_{i<j}^{2n}e^{\sigma_i\sigma_j F_2(t_i-t_j)},\qquad \qquad {\rm for} \qquad \sigma_i=\pm 1,\qquad {\rm and} \qquad \sum_{i=1}^{2n} \sigma_i=0.
\end{equation}
Using this decomposition the expression Eq.~\eqref{suppeq:p2full} can be represented diagrammatically as shown in Fig.~(\ref{suppfig:suppFig1}). 

\subsection{Decorrelation approximation}
For further approximations and also for numerical integration, the above integral can be simplified by symmetrizing  the integration domains and written in a compact form as  
\begin{equation}\label{eq:p2_SingleTerm}
\begin{split}
p_2(t\rightarrow \infty) 
= & 
\frac{\eta^4}{8}\int_0^{t\rightarrow \infty}dt_4 \int_0^{t_4}dt_2 \int_0^{t\rightarrow\infty}dt_3\int_0^{t_3}dt_1 \\
& e^{(i(2\Delta+2\Delta_p)-\kappa)(t-t_4)+(i\Delta-\frac{\kappa}{2})(t_4-t_2)}
e^{(-i(2\Delta+2\Delta_p)-\kappa)(t-t_3)+(-i\Delta-\frac{\kappa}{2})(t_3-t_1)} \\
& \langle e^{iP(t_1)}e^{iP(t_3)}e^{-iP(t_4)}e^{-iP(t_2)}\rangle_{\rm th}.
\end{split}
\end{equation}

Further, since we are interested in the steady state, we can use $\dot p_2(t\rightarrow \infty)=0$ and eliminate one of the integrals,
\begin{equation}\label{eq:p2_ThreeIntegrals}
\begin{split}
p_2(t\rightarrow \infty) 
= & 
\frac{\eta^4}{8\kappa} {\rm Re} \int_0^{\infty} dt_3 \int_0^{\infty}d t_2 \int_0^{t_3} d t_1 \\
& e^{(-i(2\Delta+2\Delta_p)-\kappa)(t-t_3)} e^{(i\Delta-\frac{\kappa}{2})(t-t_2)}
e^{(-i\Delta-\frac{\kappa}{2})(t_3-t_1)} \,  \langle e^{iP(t_1)}e^{iP(t_3)}e^{-iP(t)}e^{-iP(t_2)}\rangle_{\rm th}.
\end{split}
\end{equation}
In Fig. \ref{fig:Fig3}(c) in the main text, this integral has been evaluated numerically using the analytic result for the four-point correlator, i.e., Eqs.~\eqref{suppeq:Wick} and~\eqref{suppeq:F2corr}.

\subsubsection{Resonant two-photon correlations}
The effect of photon blockade is most pronounced when the cavity is driven at the single photon resonance, $\Delta^{(s)}_{\rm res}\approx \Delta_p$, and for small $\kappa\ll \Delta_p$, where the second transition is then detuned by $\sim 2\Delta_p$ from the two-photon resonance. 
Under this resonance condition and assuming $\kappa \ll \omega_c$ further simplifications of the two-photon correlation function are possible. As a starting point, we use the expression for the two-photon probability given in  Eq.~\eqref{eq:p2_SingleTerm} and decompose the four-point correlation function according to Eq.~\eqref{suppeq:Wick}. For resonant driving, $\Delta\simeq 0$, we then obtain
\begin{equation}\label{suppeq:p2simpleI}
\begin{split}
p_2(t\rightarrow \infty) 
\simeq & 
\frac{\eta^4}{8}\int_0^{t\rightarrow \infty} dt_4 \int_0^{t\rightarrow\infty} d t_3\, 
\underbrace{e^{(i2\Delta_p-\kappa)(t-t_4)}
    e^{(-i2\Delta_p-\kappa)(t-t_3)}e^{-F_2(t_3 -t_4)}}_{=A(t_4,t_3)}\\
&\;\;\;  \times \;\; \underbrace{\int_0^{t_4} d t_2 \int_0^{t_3} d t_1\, 
    e^{-\frac{\kappa}{2}(t_4-t_2)}e^{-\frac{\kappa}{2}(t_3-t_1)} e^{F_2(t_1-t_3) - F_2(t_1-t_4) -F_2(t_1-t_2)-F_2(t_3-t_2)+F_2(t_4-t_2)}}_{=B(t_4,t_3)}.
\end{split}
\end{equation}
The integrand is made up of an oscillating, $A(t_4,t_3)$, and a slowly varying, $B(t_4,t_3)$, component. Therefore, we can use the approximate identity $\int_0^{t\rightarrow \infty}d t_4 \int_0^{t\rightarrow \infty}d t_3\, A(t_4,t_3)B(t_4,t_3)\simeq B(t_4\rightarrow \infty,t_3\rightarrow \infty) \int_0^{t\rightarrow \infty}dt_4\int_0^{t\rightarrow\infty}dt_3 \,A(t_4,t_3)$ for $\partial_{t_i}B(t_4,t_3)\ll\partial_{t_i}A(t_4,t_3) $ and obtain
\begin{equation}\label{eq:p2simpleII}
\begin{split}
p_2(t\rightarrow \infty) 
\simeq & 
\frac{\eta^4}{8}\int_0^{t\rightarrow \infty}dt_4\int_0^{t\rightarrow\infty}dt_3\, 
e^{(i2\Delta_p-\kappa)(t-t_4)}
e^{(-i2\Delta_p-\kappa)(t-t_3)}e^{-F_2(t_3 -t_4)}\\
& \;\; \times\,\, \int_0^{t_4\rightarrow \infty }dt_2 \int_0^{t_3\rightarrow \infty}dt_1
e^{-\frac{\kappa}{2}(t_4-t_2)}e^{-\frac{\kappa}{2}(t_3-t_1)}e^{-F_2(t_1-t_2)} \\
& \;\;\;\;\;\;\;\;\;\;\;\;\;\;\;\;\;\;\;\;\;\;\;\;\;\;\;\times \,\, \underbrace{\left(e^{F_2(t_1-t_3) - F_2(t_1-t_4) -F_2(t_3-t_2)+F_2(t_4-t_2)}\right)\Bigr\vert_{t_{3,4}\rightarrow \infty} }_{=: X}.
\end{split}
\end{equation}
In a final step we use that in the limit $t_3\sim t_4 \gg t_1,t_2$ the four point correlator, which appears in Eq.~\eqref{eq:p2_SingleTerm}, can be approximated by  
\begin{equation}
\begin{split}
-F_{4}(\{ t_j\})=&F_2(t_1-t_3)- F_2(t_1-t_4)-F_2(t_1-t_2)-F_2(t_3-t_4)- F_2(t_3-t_2)+F_2(t_4-t_2)\\
\approx &-F_2(t_1-t_2)-F_2(t_3-t_4).
\end{split}
\end{equation}

This approximation is justified by the fact that the function $F_2(\tau)$ varies slowly for large times $\tau\gg \omega_c^{-1}$ and therefore all the terms with large time arguments cancel. Indeed $F(\tau\gg \omega_c^{-1}) \simeq const.$ for $s=2$ and $F(\tau\gg \omega_c^{-1})\sim \ln(\tau)$ for $s=1$. For the orginal four-point correlator, this cancellation corresponds to the factorization
\begin{equation}
\langle e^{iP(t_1)}e^{iP(t_3)} e^{-iP(t_4)}e^{-iP(t_2)}\rangle_{\rm th} \approx \langle e^{iP(t_1)}e^{-iP(t_2)}\rangle_{\rm th} \langle e^{iP(t_3)} e^{-iP(t_4)}\rangle_{\rm th},
\end{equation}
which means that the first and a possible second photon absorption process are uncorrelated. Equivalently, this approximation means that  in Eq.~\eqref{eq:p2simpleII} we can set $X\approx 1$ and the two double integrals in the first and the second line factorize. In total we obtain
\begin{equation}
\begin{split}
p_2(t\rightarrow \infty) \simeq n_0 S^{(s)}_{\alpha,2\kappa}(\Delta_p) \times n_0 S^{(s)}_{\alpha,\kappa}(-\Delta_p),
\end{split}
\end{equation}
which leads to Eq.~\eqref{eq:g20simple} from the main text. These approximations suggest the interpretation as sequential absorption of two-single photons and therefore we expect this to be only accurate as long as the quasi-photon picture is valid; for all $\alpha$ if $s=2$ and for $\alpha<1/2$ for $s=1$. 

\begin{figure}
    \includegraphics[width=0.7\textwidth]{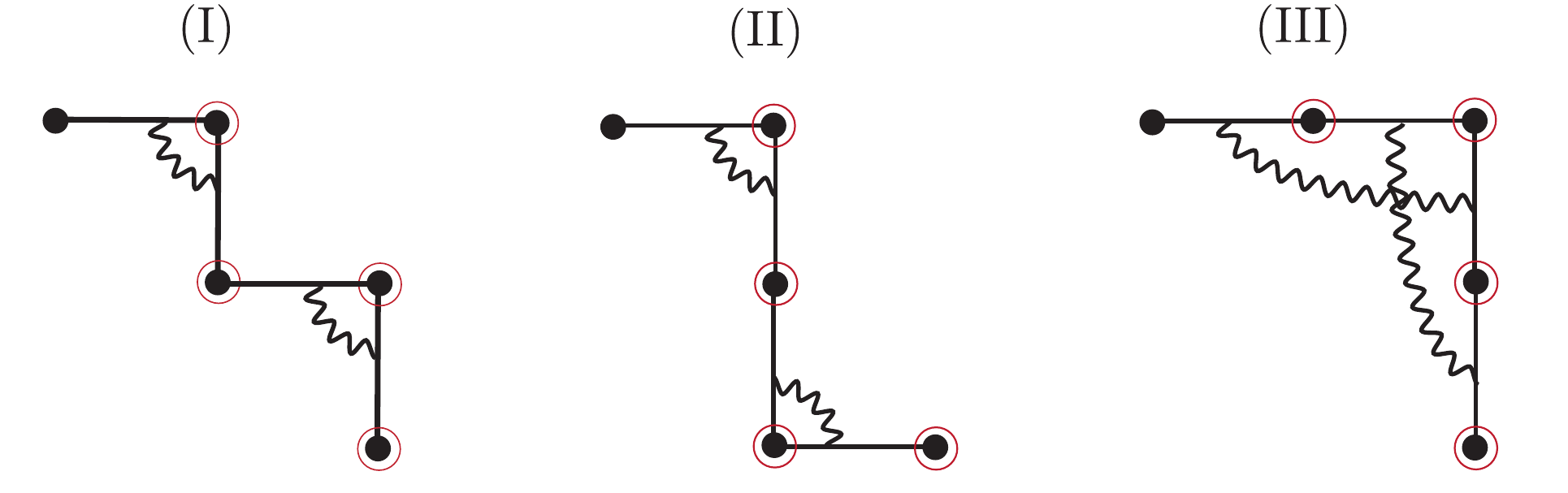} 
    \caption{Photon paths and remaining polaron correlations contributing to two-photon population in the case of single photon absorption picture corresponding to Eq.~\eqref{eq:p2pathsimple}. }
    \label{suppfig:suppFig2}
\end{figure}

It is instructive to convert this simplified result back into the form of Eq.~\eqref{suppeq:p2full}, in which case we  obtain
\begin{equation} \label{eq:p2pathsimple}
\begin{split}
p_2(t\rightarrow \infty) 
\approx   \frac{\eta^4}{8} &\int_0^{t\rightarrow \infty}dt_4\int_0^{t_4}dt_3\int_0^{t_3}dt_2\int_0^{t_2} dt_1 \\
&\Bigl[  \mathcal{G}_{22}(t-t_4)\mathcal{G}_{12}(t_4-t_3)\mathcal{G}_{11}(t_3-t_2)\mathcal{G}_{01}(t_2-t_1)\average{ e^{iP(t_1)}e^{-iP(t_2)}}_{\rm th}\average{e^{iP(t_3)} e^{-iP(t_4)}}_{\rm th}  \qquad \,\,\,\,\,({\rm I}) \\
&  + \mathcal{G}_{22}(t-t_4)\mathcal{G}_{21}(t_4-t_3)\mathcal{G}_{11}(t_3-t_2)\mathcal{G}_{01}(t_2-t_1)\average{ e^{iP(t_1)}e^{-iP(t_2)}}_{\rm th}\average{e^{iP(t_4)}e^{-iP(t_3)}}_{\rm th} \qquad ({\rm II}) \\
& + \mathcal{G}_{22}(t-t_4)\mathcal{G}_{12}(t_4-t_3)\mathcal{G}_{02}(t_3-t_2)\mathcal{G}_{01}(t_2-t_1)\average{ e^{iP(t_1)}e^{-iP(t_3)}}_{\rm th}\average{e^{iP(t_2)}e^{-iP(t_4)}}_{\rm th} \qquad ({\rm III}) \\ & +\mathrm{c.c.}\Bigr]. 
\end{split}
\end{equation}
These factorized correlations correspond to the simplified diagrams depicted in Fig.~(\ref{suppfig:suppFig2}) and show that within this decorrelation approximation all correlations in each excitation path between the first and the second photon are neglected.  

\section{Time Evolved Matrix Product Operators}

In this section we give an overview of the the time-evolving matrix product operator (TEMPO) algorithm used in the main text to give an exact numerical comparison to the pertubation theory results. The method works by  representing the Feynman-Vernon path integral in terms of a tensor network, taking fully into account the non-Markovian dynamics produced. For full details we refer the reader to Ref.~\cite{suppStrathearn2018}.

By formally integrating out the Gaussian environment it is possible to find a representation of the history of the system in terms of an augmented density tensor (ADT) which after $k$ timesteps take the form of a tensor $A^{i_1, i_2, \ldots, i_k}(t_k)$ where each index runs over the $d^2$ possible density matrix elements of a system in a $d$-dimensional Hilbert space. In this language the path integral can be written as a propagator for the ADT which adds an extra index corresponding to the new point in history which needs to be added this then takes the form $B_{i_1, i_2, \ldots i_k}^{j_1, j_2, \ldots j_{k+1}}$. Up to this point the only approximation introduced is that we have discretized time onto a fixed grid of points which gives a Trotter error of $\mathcal{O}(dt^3)$. Currently the memory requirements scale exponentially with the number of timesteps, to overcome this we use the memory cutoff approximation which means we remove the influence on the system from time points more than $K$ in the past. This then reduces the requirements to scale only exponentially in $K$ and is equivalent to the QUAPI algorithm proposed by Makri and Marakov~\cite{suppmakri_makarov_1995_i, suppmakri_makarov_1995_ii} and clearly described in~\cite{suppVagov2011, suppStrathearn2017}. The advantage gained by using TEMPO is to realise that the multiplication of the ADT, $A$ by the $B$ propagators can be written as a tensor network and then by writing $A$ as a matrix product state
\begin{equation}
A^{i_1, i_2, \ldots, i_k}(t_k) = \sum_{\alpha_1, \alpha_2, \ldots \alpha_{k}}[a^{i_1}]_{\alpha_1}[a^{i_2}]_{\alpha_1, \alpha_2}\ldots [a^{i_k}]_{\alpha_k}
\end{equation}
we may perform singular value decompositions over the $\alpha$ indices removing the states below some threshold and thus significantly reducing the memory requirements of the propagation.

\subsection{Convergence}

\begin{table}
    \begin{tabular}{|c|c|c|c|}
        \hline
        Result & Hilbert space, $\mathcal{H}$ & Memory length, $K$ & Singular value cutoff, $\chi_{\text{min}}$ \\
        \hline
        Fig.~2 & 4 & 60 & $10^{-7}$ \\
        Fig.~3 & 4 & 90 & $10^{-9}$ \\
        Fig.~4, $\alpha = 0.01$ & 8 & 10 & $10^{-4}$ \\
        Fig.~4, $\alpha = 0.05$ & 6 & 30 & $10^{-5}$ \\
        Fig.~4, $\alpha = 0.15$ & 5 & 50 & $10^{-6}$ \\
        \hline
    \end{tabular}
    \caption{\label{tab:conv} Values used to achieve numerical convergence in the results presented in the main text.}
\end{table}

To ensure convergence of the numerical results there are four parameters which need to be checked:
\begin{itemize}
    
    \item The size of the timestep used in the discretization process $dt$ 
    \item The number of states in the bosonic Hilbert space, $\mathcal{H}$
    \item The number of time points $K$ before making the memory cutoff
    \item The size of singular values (relative to the largest), $\chi_{\text{min}}$, below which they are thrown away
\end{itemize}
We ensured that none of the results in the main text change with any of these parameters for the values chosen. The result of this is that we find generally for stronger system environment coupling $\alpha$ we need to keep a longer memory length with a stricter singular value cutoff, but this is roughly compensated by the fact that the local Hilbert space can be smaller since higher photon states do not become populated.

The values required are  summarized in Table~\ref{tab:conv}. For all results we find that $dt=0.1\omega_c$ results in converged results. For Figs.~\ref{fig:Fig2} and~\ref{fig:Fig3} we used the values required for the strongest coupling over the whole range and the values required were slightly more stringent for the the two photon expectation values required to calculate $g^{(2)}(0)$. The dynamics presented in Fig.~\ref{fig:Fig4} of the main text required different parameters depending on the value of the system-environment coupling strength $\alpha$. At small $\alpha$ the memory length can be short but a large number of states are required to capture the dynamics, while at larger $\alpha$ the opposite is true.

%

\end{document}